\documentclass[pra,showpacs,preprintnumbers,amsmath,superscriptaddress,aps]{revtex4}

\def\duzomniejsze{<\kern-.7mm<}
\def\duzowieksze{>\kern-.7mm>}

\def\textbf#1{{\bf #1}}
\def\be{\begin{equation}}
\def\ee{\end{equation}}
\def\bea{\begin{eqnarray}}
\def\eea{\end{eqnarray}}
\def\bse{\begin{subequations}}
\def\ese{\end{subequations}}

\newcommand{\bei}{\begin{itemize}}
\newcommand{\eei}{\end{itemize}}
\newcommand{\bee}{\begin{enumerate}}
\newcommand{\eee}{\end{enumerate}}

\def\>{\rangle}
\def\<{\langle}

\def\dt#1{{{\kern -.0mm\rm d}}#1\,}

\newtheorem{theorem}{Theorem}
\newtheorem{definition}{Definition}

\begin{document}

\title{A simple proof of monogamy of entanglement}
\author{Dong Yang}
\affiliation{Hefei National Laboratory for Physical Sciences at Microscale \& Department of Modern Physics, University of Science and Technology of China, Hefei, Anhui 230026, Peoples Republic of China}

\date{\today}

\begin{abstract}
Monogamy of entanglement means that an entangled state cannot be shared with many parties. The more parties, the less entanglement between them. In this paper, we give a simple proof of this property and provide an upper bound of the number of parties.
\end{abstract}


\maketitle

Monogamy is one of crucial properties of entanglement. It is essential in quantum cryptography. A simple example is the Bell state $|\Phi\>_{AB}=1/\sqrt{2}(|00\>+|11\>$ shared between Alice and Bob. Monogamy of the pure entangled state $|\Phi\>_{AB}$ excludes any possibility that another party including the potential eavesdropper Eve could correlate. The monogamous property of the pure entangled state is extended to the un-sharable property for the mixed state \cite{Terhal,Werner,Doherty}. In a recent paper \cite{Bae}, the monogamous property is employed to prove asymptotic quantum cloning is state estimation, which has been identified as one of the open problems in quantum information theory. In this paper, we give a simple proof of the monogamy of entanglement and provide an upper bound of the number of parties.\\

{\definition A bipartite state $\rho_{AB}$ is said to be n-sharable when it is possible to find a quantum state $\rho_{AB_1B_2\cdots B_n}$ such that $\rho_{AB_1}=\rho_{AB_2}=\cdots=\rho_{AB_n}=\rho_{AB}$ where $ \rho_{AB_k}=tr_{B_{\bar{k}}}\rho_{AB_1B_2\cdots B_n}$. If such state exists, $\rho_{AB_1B_2\cdots B_n}$ is called as an n-extension of $\rho_{AB}$.}

{\theorem \cite{Terhal,Werner,Doherty} A bipartite state is n-sharable for any $n$ if and only if it is separable.}\\

{\it Proof.} For a separable state $\rho_{AB}$, there always exists a separable decomposition \be \rho_{AB}=\sum_{i}p_i(|\phi_i\>\<\phi_i|)_A\otimes(|\psi_i\>\<\psi_i|)_B. \ee It is explicit that \be \rho_{AB_1B_2\cdots B_n}=\sum_{i}p_i(|\phi_i\>\<\phi_i|)_A\otimes(|\psi_i\>\<\psi_i|)^{\otimes n}_{B_1B_2\cdots B_n} \ee
is a valid n-extension of $\rho_{AB}$ for any $n$.\\

Next we prove that for any entangled state $\rho_{AB}$, there always exists a finite $N$ such that no valid n-extension can be found for any $n> N$. Recall that the duality relation of a pure tripartite state $\phi_{ABC}$  is \cite {Koashi} \be S(\rho_{A})=E_{f}(\rho_{A:B})+C_{\leftarrow}(\rho_{A:C}). \ee Here $S(\rho_{A})=-tr\rho_A\log\rho_{A}$ is the von Neumann entropy of $\rho_{A}$. $E_f(\rho_{AB})=\min{\sum_ip_iE(\phi^i_{AB})}$ is the entanglement of formation (EoF) \cite{Bennett1}, where the minimum is taken over all pure ensembles $\{p_i,|\phi^i\>_{AB}\}$ satisfying $\rho_{AB}=\sum_ip_i(|\phi^i\>\<\phi^i|)_{AB}$, and entanglement for pure state $\phi_{AB}$ is $E(\phi_{AB})=S(tr_B(|\phi\>\<\phi|)_{AB})$. $C_{\leftarrow}(\rho_{A:C})=\max_{C_{i}^{\dagger}C_{i}}S(\rho_A)-\sum_{i}p_{i}S(\rho^{i}_{A})$ is the classical correlation of bipartite state $\rho_{AC}$ \cite{HV}, where $\{C_{i}^{\dagger}C_{i}\}$ is a positive operator-valued measurement (POVM) performed on subsystem $C$, $\rho^{i}_{A}=tr_C(I\otimes C_i\rho_{AC}I\otimes C_i^{\dagger} )/p_i$ is the remaining state of $A$ after obtaining the outcome $i$ on $C$, and $p_i=tr_{AC}(I\otimes C_i\rho_{AC}I\otimes C_i^{\dagger})$ is the probability to obtain outcome $i$.

Suppose the optimal decomposition of EoF for a mixed tripartite state $\rho_{A:BC}$ is $\{p_i,|\phi^i\>_{A:BC}\}$, we have \bea
E_f(\rho_{A:BC})&=&\sum p_iS(\rho_{A}^i)\\
&=&\sum p_i(E_{f}(\rho_{A:B}^i)+C_{\leftarrow}(\rho_{A:C}^i))\\
&\ge&E_f(\rho_{A:B})+G_{\leftarrow}(\rho_{A:C}). \label{inequ} \eea The inequality $(\ref{inequ})$ comes from the convexity of EoF $\sum p_iE_{f}(\rho_{A:B}^i)\ge E_{f}(\sum p_i\rho_{A:B}^i)$ \cite{Bennett1} and $G_{\leftarrow}(\rho_{A:C})=\min\sum_ip_iC_{\leftarrow}(\rho^i_{A:C})$ \cite{DM}, where the minimum is taken over all mixed ensembles $\{p_i,\rho^i_{AC}\}$ satisfying $\rho_{AC}=\sum_ip_i\rho^i_{AC}$.

Now iteratively applying the inequality $(\ref{inequ})$ to the n-extension state $\rho_{AB_1B_2\cdots B_n}$ of an entangled state $\rho_{AB}$ and further noticing $E_f(\rho_{A:B})\ge G_{\leftarrow}(\rho_{A:B})$ by definition of $G_{\leftarrow}$ \cite{DM}, we obtain \bea
E_f(\rho_{A:B_1B_2\cdots B_n})&\ge& E_f(\rho_{A:B_2\cdots B_{n}})+G_{\leftarrow}(\rho_{A:B_1})\nonumber\\
&\ge& E_f(\rho_{A:B_3\cdots B_{n}})+G_{\leftarrow}(\rho_{A:B_2})+G_{\leftarrow}(\rho_{A:B_1})\nonumber\\
&=&E_f(\rho_{A:B_3\cdots B_{n}})+2G_{\leftarrow}(\rho_{A:B})\nonumber\\
&\ge& \cdots\cdots\nonumber\\
&\ge& E_f(\rho_{A:B})+(n-1)G_{\leftarrow}(\rho_{A:B})\nonumber\\
&\ge& nG_{\leftarrow}(\rho_{A:B}). \eea Employing the explicit relation $S(\rho_A)\ge E_f(\rho_{A:B_1B_2\cdots B_n})$ and an important property of $G_{\leftarrow}$ asserting that {\it $G_{\leftarrow}(\rho_{A:B})>0$ if and only if $\rho_{AB}$ is entangled} \cite{DM}, we get the upper bound of n-extension of an entangled state $\rho_{AB}$, \be n\le N=[\frac{S(\rho_{A})}{G_{\leftarrow}(\rho_{AB})}], \ee
where $[x]$ is the maximal integer not larger than $x$. Thus we proved that for entangled state, there exists a finite number $N$ such that no n-extension can be found for $n>N$. Physically, entanglement in a given state can not be sharable in arbitrarily many parties.\\

For the system of many identical particles, the state $\rho_{A_1A_2,\cdots,A_n}$ has the symmetry of permutation. It holds for any state that entanglement between any pair particles tends to zero as $n\to\infty$. \\

In summary, we give a simple proof of monogamy of entanglement and provide an upper bound of the number of parties beyond which entanglement cannot be shared.\\

{\it Acknowledgement} D. Yang would like to thank M. Horodecki for helpful comment.

\end{document}